\documentclass[11pt]{article}
\usepackage{amsmath,amssymb,amsthm,amsxtra}
\def\thefootnote{\fnsymbol{footnote}}

\begin{document}

\vspace{0.2cm}

\begin{center}
{\large\bf A Generic Diagonalization of the $3\times 3$ Neutrino
Mass Matrix and Its Implications on the $\mu$-$\tau$ Flavor Symmetry and
Maximal CP Violation}
\end{center}

\vspace{0.1cm}

\begin{center}
{\bf Zhi-zhong Xing}
\footnote{E-mail: xingzz@ihep.ac.cn}
$~$ and $~$ {\bf Ye-Ling Zhou} \\
{\sl Institute of High Energy Physics,
Chinese Academy of Sciences, Beijing 100049, China}
\end{center}

\vspace{1.5cm}

\begin{abstract}
In the flavor basis where the mass eigenstates of three charged
leptons are identified with their flavor eigenstates, one may
diagonalize a $3\times 3$ Majorana neutrino mass matrix
$M^{}_\nu$ by means of the standard parametrization of the $3\times
3$ neutrino mixing matrix $V$. In this treatment the unphysical
phases of $M^{}_\nu$ have to be carefully factored out, unless a
special phase convention for neutrino fields is chosen so as
to simplify $M^{}_\nu$ to $M^\prime_\nu$ without any unphysical
phases. We choose this special flavor basis and establish some
exact analytical relations between the matrix elements of
$M^\prime_\nu {M^\prime_\nu}^\dagger$ and seven physical parameters
--- three neutrino masses ($m^{}_1$, $m^{}_2$, $m^{}_3$), three
flavor mixing angles ($\theta^{}_{12}$, $\theta^{}_{13}$,
$\theta^{}_{23}$) and the Dirac CP-violating phase ($\delta$). Such
results allow us to derive the conditions for the $\mu$-$\tau$
flavor symmetry with $\theta^{}_{23} = \pi/4$ and maximal CP
violation with $\delta = \pm\pi/2$, which should be useful for
discussing specific neutrino mass models. In particular, we
show that $\theta^{}_{23} =\pi/4$ and $\delta =
\pm \pi/2$ keep unchanged when constant matter effects are taken
into account for a long-baseline neutrino oscillation experiment.
\end{abstract}

\begin{center}
PACS number(s): 14.60.Pq, 13.10.+q, 25.30.Pt
\end{center}

\def\thefootnote{\arabic{footnote}}
\setcounter{footnote}{0}

\newpage

\framebox{\large\bf 1} \hspace{0.1cm} Recent neutrino oscillation
experiments \cite{PDG10} have provided us with very convincing
evidence that neutrinos are massive and lepton flavors are mixed.
Similar to the phenomenon of quark flavor mixing, which is described
by the $3\times 3$ Cabibbo-Kobayashi-Maskawa (CKM) matrix
\cite{CKM}, the phenomenon of lepton flavor mixing can also be
described by a $3\times 3$ unitary matrix $V$, the so-called
Maki-Nakagawa-Sakata-Pontecorvo (MNSP) matrix \cite{MNSP}
\footnote{Different from the CKM matrix, which must be unitary in
the standard electroweak model, the MNSP matrix $V$ can be either
unitary or non-unitary in a given neutrino mass model. For example,
$V$ is unitary in the type-II seesaw mechanism, but it must be
non-unitary in the type-I, type-(I+II), type-III and multiple seesaw
models \cite{Xing2010} although the effects of its unitarity
violation are at most at the percent level \cite{Antusch}. In this
paper we simply assume that $V$ is unitary at low energies.}.
A full parametrization of the MNSP matrix $V$ needs three rotation
matrices in the complex (1,2), (1,3) and (2,3) planes:
\begin{eqnarray}
O^{}_{12} & = & \begin{pmatrix} c^{}_{12} & s^{}_{12} & 0 \\
-s^{}_{12} & c^{}_{12} & 0 \\ 0 & 0 & 1 \end{pmatrix} \; ,
\nonumber \\
O^{}_{13} & = & \begin{pmatrix} c^{}_{13} & 0 & \hat{s}_{13}^* \\
0 & 1 & 0 \\ -\hat{s}^{}_{13} & 0 & c^{}_{13} \end{pmatrix} \; ,
\nonumber \\
O^{}_{23} & = & \begin{pmatrix} 1 & 0 & 0 \\ 0 & c^{}_{23} &
s^{}_{23}
\\ 0 & -s^{}_{23} & c^{}_{23} \end{pmatrix} \; ,
\end{eqnarray}
where $c^{}_{ij} \equiv \cos\theta^{}_{ij}$, $s^{}_{ij} \equiv
\sin\theta^{}_{ij}$ and $\hat{s}^{}_{13} \equiv s^{}_{13} \
e^{i\delta}$ (for $ij=12, 13, 23$). The unitary MNSP matrix $V$ can
then be parametrized as
\begin{equation}
V \; = \; P^{}_l U P^{}_\nu \; ,
\end{equation}
where $P^{}_l = {\rm Diag}\{e^{i\phi^{}_e}, e^{i\phi^{}_\mu},
e^{i\phi^{}_\tau} \}$ and $P^{}_\nu = {\rm Diag}\{e^{i\rho},
e^{i\sigma}, 1\}$ are two diagonal phase matrices, and
\begin{equation}
U \; = \; O^{}_{23} O^{}_{13} O^{}_{12} \; = \; \begin{pmatrix}
c^{}_{12} c^{}_{13} & s^{}_{12} c^{}_{13} & \hat{s}^*_{13} \\
-s^{}_{12} c^{}_{23} - c^{}_{12} s^{}_{23} \hat{s}^{}_{13} &
c^{}_{12} c^{}_{23} - s^{}_{12} s^{}_{23} \hat{s}^{}_{13} &
s^{}_{23} c^{}_{13} \\ s^{}_{12} s^{}_{23} - c^{}_{12} c^{}_{23}
\hat{s}^{}_{13} & -c^{}_{12} s^{}_{23} - s^{}_{12} c^{}_{23}
\hat{s}^{}_{13} & c^{}_{23} c^{}_{13} \end{pmatrix} \;
\end{equation}
is just the standard parametrization of the CKM matrix \cite{PDG10}.
Without loss of generality, one may arrange three mixing angles
($\theta^{}_{12}$, $\theta^{}_{13}$, $\theta^{}_{23}$) to lie in the
first quadrant and allow three CP-violating phases ($\delta$,
$\rho$, $\sigma$) to vary between $0$ and $2\pi$. In the mass
eigenstates of three charged leptons and three neutrinos, $V$
appears in the charged-current interactions
\begin{equation}
{\cal L}^{}_{\rm cc} \; =\; - \frac{g}{\sqrt{2}} \
\overline{\begin{pmatrix} e & \mu & \tau \end{pmatrix}^{}_{\rm L}} ~
\gamma^\mu \ V \begin{pmatrix} \nu^{}_1 \\ \nu^{}_2 \\ \nu^{}_3
\end{pmatrix}^{}_{\rm L} W^-_\mu + {\rm h.c.} \; .
\end{equation}
Hence the phase matrix $P^{}_l$ is unphysical and can be rotated
away by redefining the phases of three charge-lepton fields. If
neutrinos are the Dirac particles, the phase matrix $P^{}_\nu$ can
also be rotated away be redefining the phases of three neutrino
fields. In this case we are left with $U$ as the MNSP matrix. If
neutrinos are the Majorana particles, however, $P^{}_\nu$ is
physical because it characterizes two irremovable relative phases of
three Majorana neutrino fields \cite{Xing04}. In this case we are
left with $V^\prime = U P^{}_\nu$, which contains three mixing
angles and three CP-violating phases. We shall assume neutrinos to
be the Majorana particles throughout this paper ($U$ in the Dirac
case can simply be reproduced from $V^\prime$ in the Majorana case
by setting $P^{}_\nu = {\bf 1}$).

In the flavor basis where the mass eigenstates of three charged
leptons are identified with their flavor eigenstates, $V^\prime = U
P^{}_\nu$ links the neutrino flavor eigenstates $(\nu^{}_e,
\nu^{}_\mu, \nu^{}_\tau)$ to the neutrino mass eigenstates
$(\nu^{}_1, \nu^{}_2, \nu^{}_3)$. The effective Majorana neutrino
mass term can be written as
\begin{equation}
{\cal L}^{}_\nu \; =\; \frac{1}{2} \ \overline{\begin{pmatrix}
\nu^{}_e & \nu^{}_\mu & \nu^{}_\tau \end{pmatrix}^{}_{\rm L}} ~
M^{}_\nu \begin{pmatrix} (\nu^{}_e)^{\rm c} \\ (\nu^{}_\mu)^{\rm c} \\
(\nu^{}_\tau)^{\rm c} \end{pmatrix}^{}_{\rm R} + {\rm h.c.} \; ,
\end{equation}
where $(\nu^{}_\alpha)^{\rm c} \equiv {\cal C}
\overline{\nu^{}_\alpha}^T$ denotes the charge-conjugate counterpart
of $\nu^{}_\alpha$ (for $\alpha =e, \mu, \tau$), and $M^{}_\nu$ is a
symmetric $3 \times 3$ matrix which totally has six complex entries
or twelve real parameters:
\begin{equation}
M^{}_\nu \; =\; \begin{pmatrix} a & b & c \\ b & d & e \\ c & e & f
\end{pmatrix} \; .
\end{equation}
In the chosen basis one may obtain three neutrino masses ($m^{}_1$,
$m^{}_2$, $m^{}_3$), three neutrino mixing angles ($\theta^{}_{12}$,
$\theta^{}_{13}$, $\theta^{}_{23}$) and three CP-violating phases
($\delta$, $\rho$, $\sigma$) by diagonalizing $M^{}_\nu$. We stress
that it is not a trivial job to diagonalize $M^{}_\nu$ and derive
the generic expressions of nine physical parameters in terms of the
elements of $M^{}_\nu$. Such an attempt has been made by Aizawa and
Yasu$\rm\grave{e}$ \cite{Yasue}, but their treatment is subject to
the transformation
\footnote{Note that the notations used in Ref. \cite{Yasue} are
different from ours.}
\begin{equation}
{V^\prime}^\dagger M^{}_\nu {V^\prime}^* \; =\; \widehat{M}^{}_\nu
\; \equiv \; \begin{pmatrix} m^{}_1 & 0 & 0 \\ 0 & m^{}_2 & 0 \\ 0 &
0 & m^{}_3 \end{pmatrix} \; .
\end{equation}
In view of the fact that $V^\prime = U P^{}_\nu$ and
$\widehat{M}^{}_\nu$ totally consist of nine physical parameters but
$M^{}_\nu$ generally contains twelve free parameters, one
immediately encounters a parameter mismatching problem. This
ambiguity will be clarified in this paper.

The purpose of our work is two-fold. First, we point out that one
should use the transformation $V^\dag M^{}_\nu V^* =
\widehat{M}^{}_\nu$ instead of the one in Eq. (7) to diagonalize the
$3\times 3$ Majorana neutrino mass matrix $M^{}_\nu$. In this case
the aforementioned parameter mismatching problem does not occur,
because $V = P^{}_l V^\prime$ contains three unphysical phases which
can exactly eliminate the unphysical phases of $M^{}_\nu$. Second,
we choose a special flavor basis of three neutrino fields to factor
out the unphysical phases of $M^{}_\nu$ such that $M^{}_\nu$ is
reduced to $M^\prime_\nu$ which only contains nine free parameters.
Then we diagonalize the Hermitian matrix $H^\prime_\nu \equiv
M^\prime_\nu {M^\prime}^\dagger_\nu$ via the transformation
$U^\dagger H^\prime_\nu U = \widehat{M}^2_\nu$. Different from Ref.
\cite{Yasue}, here a much simpler and more transparent way is found
to establish some exact analytical relations between the physical
parameters of $U$ and $\widehat{M}^{}_\nu$ and the matrix elements
of $H^\prime_\nu$. Our results can be used to work out the
conditions for the $\mu$-$\tau$ flavor symmetry with $\theta^{}_{23}
= \pi/4$ and maximal CP violation with $\delta = \pm \pi/2$.
Their usefulness is illustrated by taking a simple example of $M^{}_\nu$.
In particular, we show that $\theta^{}_{23} =\pi/4$ and $\delta =
\pm \pi/2$ keep unchanged when constant matter effects are taken
into account for a long-baseline neutrino oscillation experiment.

\vspace{0.4cm}

\framebox{\large\bf 2} \hspace{0.1cm} Let us diagonalize $M^{}_\nu$
by means of the transformation $V^\dagger M^{}_\nu V^* =
\widehat{M}^{}_\nu$. Namely, $M^{}_\nu$ can be parametrized as
follows:
\begin{equation}
M^{}_\nu \; =\; V\widehat{M}^{}_\nu V^T \; =\; P^{}_l V^\prime
\widehat{M}^{}_\nu {V^\prime}^T P^T_l = P^{}_l M^\prime_\nu P^T_l \; ,
\end{equation}
where $V = P^{}_l V^\prime$ with $V^\prime = UP^{}_\nu$, and
$M^\prime_\nu \equiv V^\prime \widehat{M}^{}_\nu {V^\prime}^T$.
Since $U$, $P^{}_\nu$ and $\widehat{M}^{}_\nu$ contain four, two and
three real parameters respectively, $M^\prime_\nu$ totally consists
of nine parameters which are all physical or experimentally
observable. Given three unphysical phases in $P^{}_l$, the total
number of real parameters of $M^{}_\nu$ is therefore twelve. In
other words, $M^{}_\nu$ can be reduced to $M^\prime_\nu$ after its
three unphysical phases are factored out. This parameter counting is
certainly consistent with Eq. (6), in which six independent elements
of $M^{}_\nu$ are totally composed of twelve real parameters. After
substituting Eq. (8) into Eq. (5), we obtain
\begin{equation}
{\cal L}^{}_\nu \; =\; \frac{1}{2} \ \overline{\begin{pmatrix}
\nu^\prime_e & \nu^\prime_\mu & \nu^\prime_\tau
\end{pmatrix}^{}_{\rm L}} ~
M^\prime_\nu \begin{pmatrix} (\nu^\prime_e)^{\rm c} \\
(\nu^\prime_\mu)^{\rm c} \\
(\nu^\prime_\tau)^{\rm c} \end{pmatrix}^{}_{\rm R} + {\rm h.c.} \; ,
\end{equation}
where $\nu^\prime_\alpha = \nu^{}_\alpha \ e^{-i\phi^{}_\alpha}$
(for $\alpha = e, \mu, \tau$). In this new basis of three neutrino
fields, the corresponding Majorana neutrino mass matrix can be
parametrized as
\begin{equation}
M^\prime_\nu \; =\; V^\prime \widehat{M}^{}_\nu {V^\prime}^T \; =\;
U P^{}_\nu \widehat{M}^{}_\nu P^T_\nu U^T \; .
\end{equation}
Now it becomes obvious that the treatment in Ref. \cite{Yasue} is
equivalent to a choice of the special flavor basis given in Eq. (9).
This observation does not change even if one considers the following
Hermitian matrices:
\begin{eqnarray}
H^\prime_\nu & \equiv & M^\prime_\nu {M^\prime_\nu}^\dag \; = \;
V^\prime \widehat{M}_\nu^2 {V^\prime}^\dag \; = \; U P^{}_\nu
\widehat{M}_\nu^2 P^\dag_\nu U^\dag \; = \; U \widehat{M}_\nu^2
U^\dag \; ,
\nonumber \\
H^{}_\nu & \equiv & M^{}_\nu M_\nu^\dag \; = \; V \widehat{M}_\nu^2
V^\dag \; = \; P^{}_l V^\prime \widehat{M}_\nu^2 {V^\prime}^\dag
P^\dag_l \; =\; P^{}_l H^\prime_\nu P^\dagger_l \; .
\end{eqnarray}
Two comments are in order.
\begin{itemize}
\item     $H^\prime_\nu$ only contains seven real parameters and
has nothing to do with the Majorana phase
matrix $P^{}_\nu$. Hence one may establish the direct relations
between the elements of $H^\prime_\nu$ and the physical parameters
of $U$ and $\widehat{M}^{}_\nu$ (i.e., $m^{}_1$, $m^{}_2$, $m^{}_3$;
$\theta^{}_{12}$, $\theta^{}_{13}$, $\theta^{}_{23}$ and $\delta$),
no matter whether massive neutrinos are the Dirac or Majorana particles.

\item     To diagonalize $H^{}_\nu$ or $M^{}_\nu$ itself, one has to
take into account the unphysical phase matrix $P^{}_l$. In the
literature many authors have chosen the flavor basis defined in Eq.
(9) to reconstruct the effective Majorana neutrino mass matrix
$M^\prime_\nu$. This special phase convention or basis choice is
useful in the study of neutrino phenomenology, but one should keep
in mind that a neutrino mass model generally predicts $M^{}_\nu$ in
the flavor basis defined in Eq. (5).
\end{itemize}
After clarifying the difference between the flavor bases associated
with $M^{}_\nu$ (or $H^{}_\nu$) and $M^\prime_\nu$ (or
$H^\prime_\nu$), we shall follow a phenomenological way to derive
the exact analytical expressions of three neutrino masses, three
flavor mixing angles and the Dirac CP-violating phase in terms of
the matrix elements of $H^\prime_\nu$. Our present treatment is much
simpler and more transparent than the one given Ref. \cite{Yasue},
and it leads us to two constraint equations for the matrix elements
of $H^\prime_\nu$ which were not presented in Ref. \cite{Yasue}.

To be more specific, we denote the matrix elements of the Hermitian
matrix $H^\prime_\nu$ as
\begin{equation}
H^\prime_\nu \; =\; \begin{pmatrix} A & B & C \\ B^* & D & E \\
C^* & E^* & F \end{pmatrix} \; ,
\end{equation}
where $A$, $D$ and $F$ are real, and $B$, $C$ and $E$ are in general
complex. Note that $H^\prime_\nu = P^\dag_l H^{}_\nu P_l = P^\dag_l
M^{}_\nu M^\dag_\nu P_l$ holds, and the matrix elements of
$M^{}_\nu$ have been expressed in Eq. (6). Therefore,
\begin{eqnarray}
A & = & |a|^2 + |b|^2 + |c|^2 \; ,
\nonumber \\
B & = & \left( ab^* + bd^* + c e^* \right) e^{i ( \phi^{}_\mu
- \phi^{}_e)} \; ,
\nonumber \\
C & = & \left( ac^* + be^* + cf^* \right) e^{i ( \phi^{}_\tau
- \phi^{}_e)} \; ,
\nonumber \\
D & = & |b|^2 + |d|^2 + |e|^2 \; ,
\nonumber \\
E & = & \left( bc^* + de^* + ef^* \right) e^{i ( \phi^{}_\tau
- \phi^{}_\mu)} \; ,
\nonumber \\
F & = & |c|^2 + |e|^2 + |f|^2 \; .
\end{eqnarray}
Let us reiterate that the above six matrix elements of
$H^\prime_\nu$ are not fully independent. Because $H^\prime_\nu$
totally consists of seven real parameters, there must exist two
constraint equations among $A$, $B$, $C$, $D$, $E$ and $F$. This
point can also be understood in another way. The Majorana phases of
$M^{}_\nu$ (i.e., $\rho$ and $\sigma$) are completely canceled in
the elements of $H^\prime_\nu$, and the unphysical phases of
$M^{}_\nu$ (i.e., $\phi^{}_e$, $\phi^{}_\mu$ and $\phi^{}_\tau$) are
also canceled in those elements. So the number of real parameters of
$H^\prime_\nu$ is not nine but seven, leading to two correlative
equations of its six matrix elements. Now we substitute $U =
O^{}_{23} O^{}_{13} O^{}_{12}$ into the expression of $H^\prime_\nu$
in Eq. (11). Then we arrive at
\begin{equation}
O_{23}^\dag H^\prime_\nu O^{}_{23} \; =\; O^{}_{13} O^{}_{12}
\widehat{M}_\nu^2 O_{12}^\dag O_{13}^\dag \; =\; O^{}_{13} N^{}_\nu
O_{13}^\dag \; ,
\end{equation}
where
\begin{equation}
N^{}_\nu \; \equiv \; O^{}_{12} \widehat{M}_\nu^2 O^\dag_{12} \; =\;
\begin{pmatrix} m_1^2 + s_{12}^2 \Delta m_{21}^2 & c^{}_{12}
s^{}_{12} \Delta m_{21}^2 & 0 \\ c^{}_{12} s^{}_{12} \Delta m_{21}^2
& m_1^2 + c_{12}^2 \Delta m_{21}^2 & 0 \\ 0 & 0 & m_3^2
\end{pmatrix}
\; \equiv \; \begin{pmatrix} N^{}_{11} & N^{}_{12} & 0 \\
N^{}_{12} & N^{}_{22} & 0 \\ 0 & 0 & N^{}_{33} \end{pmatrix} \;
\end{equation}
with $\Delta m^2_{21} \equiv m^2_2 - m^2_1$. The left- and right-hand
sides of Eq. (14) explicitly read
\begin{eqnarray}
O_{23}^\dag H^\prime_\nu O^{}_{23} \hspace{-0.2cm} & = & \hspace{-0.3cm}
\begin{pmatrix} A & c^{}_{23} B - s^{}_{23} C & s^{}_{23} B + c^{}_{23} C
\\
c^{}_{23} B^* - s^{}_{23} C^* & c_{23}^2 D + s_{23}^2 F -
2c^{}_{23} s^{}_{23} {\rm Re}(E) & c^{}_{23} s^{}_{23} \left(D
- F\right) + c_{23}^2 E - s_{23}^2 E^* \\
s^{}_{23} B^* + c^{}_{23} C^* & c^{}_{23} s^{}_{23} \left(D -
F\right) + c_{23}^2 E^* - s_{23}^2 E & s_{23}^2 D + c_{23}^2 F +
2c^{}_{23} s^{}_{23} {\rm Re}(E) \end{pmatrix} \nonumber \\
O^{}_{13} N^{}_\nu O_{13}^\dag \hspace{-0.2cm} & = & \hspace{-0.3cm}
\begin{pmatrix} c_{13}^2 N^{}_{11} + s_{13}^2 N^{}_{33} &
c^{}_{13} N^{}_{12} & c^{}_{13} \hat{s}_{13}^*
\left(N^{}_{33} - N^{}_{11} \right) \\
c^{}_{13} N^{}_{12} & N^{}_{22} & -\hat{s}_{13}^* N^{}_{12} \\
c^{}_{13} \hat{s}^{}_{13} \left(N^{}_{33} - N^{}_{11} \right) &
-\hat{s}^{}_{13} N^{}_{12} & c_{13}^2 N^{}_{33} + s_{13}^2 N^{}_{11}
\end{pmatrix} \; .
\end{eqnarray}
The equality $(O_{23}^\dag H^\prime_\nu O^{}_{23})^{}_{12} =
(O^{}_{13} N^{}_\nu O_{13}^\dag)^{}_{12}$ yields
\begin{equation}
c^{}_{13} N^{}_{12} \; = \; c^{} _{23} B - s^{}_{23} C \; .
\end{equation}
Since the left-hand side of Eq. (17) is real and positive, we
immediately obtain ${\rm Im}(c^{}_{23} B - s^{}_{23} C) =0$. As a
result, the neutrino mixing angle $\theta^{}_{23}$ is simply given
by
\begin{equation}
\tan\theta^{}_{23} \; =\; \frac{{\rm Im}(B)}{{\rm Im}(C)} \; .
\end{equation}
On the other hand, the equality $(O_{23}^\dag H^\prime_\nu
O^{}_{23})^{}_{13} = (O^{}_{13} N^{}_\nu O_{13}^\dag)^{}_{13}$
yields
\begin{equation}
c^{}_{13} \hat{s}_{13}^* \left(N^{}_{33} - N^{}_{11} \right) \; =\;
s^{}_{23} B + c^{}_{23} C \; .
\end{equation}
This equation allows us to obtain the Dirac CP-violating phase:
\begin{equation}
\tan\delta \; = -\frac{s^{}_{23} {\rm Im}(B) + c^{}_{23} {\rm
Im}(C)}{s^{}_{23} {\rm Re}(B) + c^{}_{23} {\rm Re}(C)}
\; = \;
-\frac{\left[{\rm Im}(B)\right]^2 + \left[{\rm Im}(C)\right]^2}{{\rm
Re}(B) {\rm Im}(B) + {\rm Re}(C) {\rm Im}(C)} \; .
\end{equation}
With the help of the equality
$(O_{23}^\dag H^\prime_\nu O^{}_{23})^{}_{23} =
(O^{}_{13} N^{}_\nu O_{13}^\dag)^{}_{23}$, we have
\begin{equation}
-\hat{s}_{13}^* N^{}_{12} \; =\; c^{}_{23} s^{}_{23}
\left(D - F\right) + c_{23}^2 E - s_{23}^2 E^* \; ,
\end{equation}
which can also lead us to an expression of $\delta$:
\begin{eqnarray}
\tan\delta & = & -\frac{{\rm Im} (E)}{c^{}_{23} s^{}_{23}
\left(D - F\right) + \left(c_{23}^2 - s_{23}^2\right) {\rm Re}(E)}
\nonumber \\
& = & -\frac{{\rm Im}(E) \left\{\left[{\rm Im}(B)\right]^2 +
\left[{\rm Im}(C)\right]^2\right\}}{{\rm Im}(B) {\rm Im}(C)
\left(D - F\right) - \left\{\left[{\rm Im}(B)\right]^2 -
\left[{\rm Im}(C)\right]^2\right\}
{\rm Re}(E)} \; .
\end{eqnarray}
A straightforward comparison between Eqs. (20) and (22) yields a
constraint equation for the matrix elements of $H^\prime_\nu$:
\begin{equation}
D - F \; = \; \frac{\left\{ {\rm Re}(B) {\rm Im}(B) + {\rm Re}(C)
{\rm Im}(C) \right\} {\rm Im}(E)+ \left\{\left[{\rm Im}(B)\right]^2
- \left[{\rm Im}(C)\right]^2\right\} {\rm Re}(E)}{{\rm Im}(B) {\rm
Im}(C)} \; .
\end{equation}
Combining Eqs. (17) and (21), we obtain the smallest neutrino mixing
angle $\theta^{}_{13}$ as follows:
\begin{eqnarray}
\tan\theta^{}_{13} & = & \left| \frac{c^{}_{23} s^{}_{23} \left( D -
F \right) + c^2_{23} E - s^2_{23} E^*}{c^{}_{23} B - s^{}_{23} C}
\right| \nonumber \\
& = & \left| {\rm Im}(E) \right| \frac{\sqrt{ \left\{\left[{\rm
Im}(B)\right]^2 + \left[{\rm Im}(C)\right]^2\right\}^2 + \left\{{\rm
Re}(B) {\rm Im}(B) + {\rm Re}(C) {\rm Im}(C)\right\}^2}}{\sqrt{
\left\{ \left[{\rm Im}(B)\right]^2 + \left[{\rm Im}(C)\right]^2
\right\} \left\{ {\rm Re}(B) {\rm Im}(C) - {\rm Im}(B) {\rm Re}(C)
\right\}^2}} \; .
\end{eqnarray}
It should be noted that $\theta^{}_{13}$ can also be derived in
another way. The difference between the equalities $(O_{23}^\dag
H^\prime_\nu O^{}_{23})^{}_{11} = (O^{}_{13} N^{}_\nu
O_{13}^\dag)^{}_{11}$ and
 $(O_{23}^\dag H^\prime_\nu
O^{}_{23})^{}_{33} = (O^{}_{13} N^{}_\nu O_{13}^\dag)^{}_{33}$ reads
\begin{equation}
\left(c_{13}^2 - s_{13}^2 \right) \left(N^{}_{33} - N^{}_{11}
\right ) \; = \; s^2_{23} D + c^2_{23} F + 2c^{}_{23} s^{}_{23}
{\rm Re}(E) - A \; .
\end{equation}
Combining Eqs. (19) and (25), we obtain
\begin{eqnarray}
\tan 2\theta^{}_{13} & = & \left| \frac{2 \left( s^{}_{23} B +
c^{}_{23} C \right)}{s^2_{23} D + c^2_{23} F + 2c^{}_{23} s^{}_{23}
{\rm Re}(E) - A} \right|
\nonumber \\
& = & \frac{2 \left| B {\rm Im}(B) + C {\rm Im}(C) \right|
\sqrt{\left[{\rm Im}(B)\right]^2 + \left[{\rm Im}(C)\right]^2}}
{\left| \left( D - A \right) \left[ {\rm Im}(B) \right]^2 +
\left( F - A \right) \left[ {\rm Im}(C) \right]^2 +
2 {\rm Im}(B) {\rm Im}(C) {\rm Re}(E) \right|} \; .
\end{eqnarray}
In view of $\tan 2\theta^{}_{13} =
2\tan\theta^{}_{13}/(1-\tan^2\theta^{}_{13})$, one may do a
straightforward but lengthy calculation to work out another
constraint equation for the elements of $H^\prime_\nu$ from Eqs.
(24) and (26). The result is
\begin{eqnarray}
A & = &\frac{{\rm Re}(B) {\rm Im}(C) - {\rm Im}(B) {\rm Re}(C)}{{\rm
Im}(E)} + \frac{\left[{\rm Im}(B)\right]^2 D +\left[{\rm
Im}(C)\right]^2 F + 2{\rm Im}(B){\rm Im}(C){\rm Re}(E)}{\left[{\rm
Im}(B)\right]^2 +
\left[{\rm Im}(C)\right]^2}
\nonumber\\
& & - \frac{\left\{\left[{\rm Im}(B)\right]^2 + \left[{\rm
Im}(C)\right]^2\right\}^2 + \left\{{\rm Re}(B) {\rm Im}(B) + {\rm
Re}(C) {\rm Im}(C)\right\}^2}{\left\{\left[{\rm Im}(B)\right]^2 +
\left[{\rm Im}(C)\right]^2\right\}\left\{ {\rm Re}(B) {\rm Im}(C) -
{\rm Im}(B) {\rm Re}(C)\right\} } {\rm Im}(E) \; .
\end{eqnarray}
Eqs. (23) and (27) clearly reflect the fact that $H^\prime_\nu$ only
contains seven independent real parameters.

We proceed to derive the expressions of $\theta^{}_{12}$ and $m^2_i$
(for $i=1,2,3$) by using Eq. (15). To do so, we have to first
express $N^{}_{11}$, $N^{}_{12}$, $N^{}_{22}$ and $N^{}_{33}$ in
terms of the matrix elements of $H^\prime_\nu$. These four
quantities can be derived from Eq. (16) with the help of two
constraint equations and the results of $\theta^{}_{13}$,
$\theta^{}_{23}$ and $\delta$ obtained above. After an algebraic
exercise, we find
\begin{eqnarray}
N^{}_{11} \hspace{-0.1cm} & = & \hspace{-0.1cm} A - \frac{{\rm
Re}(B) {\rm Im}(C) - {\rm Im}(B) {\rm Re}(C)}  {{\rm Im}(E)} \; ,
\nonumber \\
N^{}_{12} \hspace{-0.1cm} & = & \hspace{-0.1cm}
\left[\frac{\left[{\rm Re}(B){\rm Im}(C)-{\rm Im}(B) {\rm Re}(C)
\right]^2}{\left[{\rm Im}(B)\right]^2 + \left[{\rm Im}(C)\right]^2}
+ \left[ \frac{\left\{{\rm Re}(B) {\rm Im}(B) + {\rm Re}(C) {\rm
Im}(C)\right\}^2}{\left\{\left[{\rm Im}(B)\right]^2 + \left[{\rm
Im}(C)\right]^2\right\}^2} + 1 \right] \left[{\rm Im}(E)\right]^2
\right]^{1/2} \hspace{-0.3cm} ,
\nonumber \\
N^{}_{22} \hspace{-0.1cm} & = & \hspace{-0.1cm} \frac{\left[{\rm
Im}(C)\right]^2 D + \left[{\rm Im}(B)\right]^2 F- 2{\rm Im}(B){\rm
Im}(C){\rm Re}(E)}{\left[{\rm Im}(B)\right]^2 + \left[{\rm
Im}(C)\right]^2} \; ,
\nonumber \\
N^{}_{33} \hspace{-0.1cm} & = & \hspace{-0.1cm} \frac{\left[{\rm
Im}(B)\right]^2 D +\left[{\rm Im}(C)\right]^2 F + 2{\rm Im}(B){\rm
Im}(C){\rm Re}(E)}{\left[{\rm Im}(B)\right]^2 + \left[{\rm
Im}(C)\right]^2} + \frac{{\rm Re}(B) {\rm Im}(C) - {\rm Im}(B) {\rm
Re}(C)} {{\rm Im}(E)} \; .
\end{eqnarray}
Then Eq. (15) leads us to
\begin{equation}
\tan2\theta_{12} \; = \; \frac{2c_{12}s_{12}}{c_{12}^2-s_{12}^2}
\; = \; \frac{2N^{}_{12}}{N^{}_{22} - N^{}_{11}} \; ,
\end{equation}
which can be expressed in terms of the elements of $H^\prime_\nu$
via Eq. (28). Furthermore, three neutrino masses can simply
be obtained from
\begin{eqnarray}
m_1^2 & = & \frac{1}{2} \left(N^{}_{11} + N^{}_{22}\right)
- \frac{1}{2} \sqrt{\left(N^{}_{22} - N^{}_{11}\right)^2 + 4N_{12}^2} \; ,
\nonumber \\
m_2^2 & = & \frac{1}{2} \left(N^{}_{11} + N^{}_{22}\right)
+ \frac{1}{2} \sqrt{\left(N^{}_{22} - N^{}_{11}\right)^2 + 4N_{12}^2} \; ,
\nonumber \\
m_3^2 & = & N^{}_{33} \; .
\end{eqnarray}
So we complete the derivation of two constraint equations for the
elements of $H^\prime_\nu$ and their exact relations with seven
physical quantities $m^2_i$ (for $i=1,2,3$), $\theta^{}_{ij}$ (for
$ij=12,13,23$) and $\delta$.

\vspace{0.4cm}

\framebox{\large\bf 3} \hspace{0.1cm} Now we consider an especially
interesting case of neutrino mixing and CP violation:
$\theta^{}_{23} =\pi/4$ and $\delta =\pm \pi/2$. In fact,
$\theta^{}_{23}$ corresponds to the $\mu$-$\tau$ flavor symmetry in
the neutrino sector in the chosen flavor basis (e.g., $|V^{}_{\mu
i}| = |V^{}_{\tau i}|$ holds (for $i=1,2,3$) in this case
\cite{Zhou08}); and $\delta =\pm \pi/2$ implies the maximal strength
of CP violation in neutrino oscillations for given values of
$\theta^{}_{12}$, $\theta^{}_{13}$ and $\theta^{}_{23}$ (i.e., the
leptonic Jarlskog parameter is maximal in this case \cite{Xing09}).
From Eqs. (18) and (20), we see that $\theta^{}_{23} =\pi/4$ and
$\delta =\pm \pi/2$ lead us to
\begin{equation}
{\rm Im}(B) \; =\; {\rm Im}(C) \; , \hspace{1cm}
{\rm Re}(B) \; =\; -{\rm Re}(C) \; ;
\end{equation}
or equivalently $B = - C^*$. In this case Eqs. (23) and (27) are
simplified to $D =F$ and
\begin{equation}
A \; =\; D + {\rm Re}(E) + 2 \frac{{\rm Re}(B) {\rm Im}(B)}{{\rm Im}(E)}
- \frac{{\rm Im}(B) {\rm Im}(E)}{{\rm Re}(B)} \; .
\end{equation}
As a consequence,
\begin{eqnarray}
\tan\theta^{}_{13} & = & \frac{1}{\sqrt{2}} \left| \frac{{\rm Im}(E)}
{{\rm Re}(B)} \right| \; ,
\nonumber \\
\tan 2\theta^{}_{12} & = & 2\frac{|{\rm Re}(B)| \sqrt{2 \left[{\rm Re}(B)
\right]^2 + \left[{\rm Im}(E)\right]^2}}
{\left| {\rm Im}(B) {\rm Im}(E) - 2 {\rm Re}(B) {\rm Re}(E) \right|}
\; ;
\end{eqnarray}
and
\begin{eqnarray}
m^2_1 & = & \frac{1}{2} \left[ 2 D - \frac{{\rm Im}(B) {\rm Im}(E)}
{{\rm Re}(B)} - \sqrt{\left[ \frac{{\rm Im}(B) {\rm Im}(E)}
{{\rm Re}(B)} - 2 {\rm Re}(E) \right]^2 + 4 \left\{ 2
\left[{\rm Re}(B)\right]^2 + \left[{\rm Im}(E)\right]^2 \right\} }
\right] \; ,
\nonumber \\
m^2_2 & = & \frac{1}{2} \left[ 2 D - \frac{{\rm Im}(B) {\rm Im}(E)}
{{\rm Re}(B)} + \sqrt{\left[ \frac{{\rm Im}(B) {\rm Im}(E)}
{{\rm Re}(B)} - 2 {\rm Re}(E) \right]^2 + 4 \left\{ 2
\left[{\rm Re}(B)\right]^2 + \left[{\rm Im}(E)\right]^2 \right\} }
\right] \; ,
\nonumber \\
m^2_3 & = & A - \frac{{\rm Im}(B) {\rm Im}(E)}
{{\rm Re}(B)} \; .
\end{eqnarray}
These simplified results are expected to be useful in discussing a
specific neutrino mass model with the $\mu$-$\tau$ flavor symmetry
and maximal CP violation.

Given the $\mu$-$\tau$ flavor symmetry and maximal CP violation, the
form of $H^\prime_\nu$ explicitly reads
\begin{equation}
{\cal H}^\prime_\nu \; =\; \begin{pmatrix} A & B & -B^* \\ B^* & D & E \\
-B & E^* & D \end{pmatrix} \; ,
\end{equation}
in which $A$, $B$ $D$ and $E$ are related to one another through Eq.
(32). Hence ${\cal H}^\prime_\nu$ totally contains five real and
independent parameters. The corresponding form of $H^{}_\nu$ defined
in Eq. (11) is
\begin{equation}
{\cal H}^{}_\nu \; =\; P^{}_l {\cal H}^\prime_\nu P^\dag_l \; =\;
\begin{pmatrix} A & B e^{i\left(\phi^{}_e - \phi^{}_\mu \right)}
& -B^* e^{i\left(\phi^{}_e - \phi^{}_\tau \right)} \\
B^* e^{i\left(\phi^{}_\mu - \phi^{}_e \right)} & D &
E e^{i\left(\phi^{}_\mu - \phi^{}_\tau \right)} \\
-B e^{i\left(\phi^{}_\tau - \phi^{}_e \right)} & E^*
e^{i\left(\phi^{}_\tau - \phi^{}_\mu \right)} & D
\end{pmatrix} \; .
\end{equation}
The meaning of this matrix is clear: if the texture of a Majorana
neutrino mass matrix ${\cal M}^{}_\nu$ derived from a specific
neutrino mass model satisfies ${\cal M}^{}_\nu {\cal M}^\dag_\nu =
{\cal H}^{}_\nu$ as given in Eq. (36), then it must predict
$\theta^{}_{23} =\pi/4$ and $\delta =\pm \pi/2$ in the standard
parametrization of the MNSP matrix. This prediction is independent
of the constraint equation in Eq. (32). Because three unphysical
phases $\phi^{}_\alpha$ (for $\alpha = e,\mu,\tau$) can be
arbitrarily rearranged, one may simply compare a model-dependent
texture of ${\cal H}^{}_\nu$ with Eq. (36) to judge whether they are
consistent with each other.

To illustrate, let us consider a typical texture of the effective
Majorana neutrino mass matrix:
\begin{equation}
{\cal M}^{}_\nu \; = \; \begin{pmatrix} a & b & -b^*
\\ b & d & e \\ -b^* & e & d^* \end{pmatrix} \;
\end{equation}
with $a$ and $e$ being real, which has been discussed in a number of
neutrino mass models with discrete flavor symmetries \cite{Example}.
Therefore,
\begin{equation}
{\cal H}^{}_\nu \; =\; {\cal M}^{}_\nu {\cal M}^\dag_\nu \; =\;
\begin{pmatrix} {\cal A} & {\cal B} & -{\cal B}^* \\
{\cal B}^* & {\cal D} & {\cal E} \\
-{\cal B} & {\cal E}^* & {\cal D}
\end{pmatrix} \; ,
\end{equation}
where
\begin{eqnarray}
{\cal A} & = & a^2 + 2|b|^2 \; ,
\nonumber \\
{\cal B} & = & \left(a - e\right) b^* + b d^* \; ,
\nonumber \\
{\cal D} & = & |b|^2 + |d|^2 + e^2 \; ,
\nonumber \\
{\cal E} & = & -b^2 + 2d e \; .
\end{eqnarray}
Comparing Eq. (38) with Eq. (36), we immediately see that they are
consistent with each other if
\begin{equation}
2\phi^{}_e \; = \; \phi^{}_\mu + \phi^{}_\tau \;
\end{equation}
is taken. In this case ${\cal H}^{}_\nu$ in Eq. (36) totally
contains six real and independent parameters: five of them come from
${\cal H}^\prime_\nu$ given in Eq. (35), and the left is just
$(\phi^{}_\mu - \phi^{}_\tau)$ because $\phi^{}_e - \phi^{}_\mu =
-(\phi^{}_\mu - \phi^{}_\tau)/2$ and $\phi^{}_e - \phi^{}_\tau =
(\phi^{}_\mu - \phi^{}_\tau)/2$ hold. In comparison, ${\cal
H}^{}_\nu$ in Eq. (38) consists of six real and independent
parameters too. That is why it is improper to take $\phi^{}_e =
\phi^{}_\mu = \phi^{}_\tau$ (or to simply assume all of them to
vanish \cite{Yasue}) and then equalize Eqs. (36) and (38).
Otherwise, the resultant parameter mismatching problem would violate
Eq. (32) and make the results in Eqs. (33) and (34) invalid. We
stress that Eq. (40) is the proper phase convention which allows us
to reduce ${\cal H}^{}_\nu$ to ${\cal H}^\prime_\nu$ in Eq. (35)
after the transformation ${\cal H}^\prime_\nu = P^\dagger_l {\cal
H}^{}_\nu P^{}_l$. Hence we must be able to arrive at the
$\mu$-$\tau$ flavor symmetry with $\theta^{}_{23} = \pi/4$ and
maximal CP violation with $\delta =\pm \pi/2$. This example clearly
shows that ${\cal H}^{}_\nu$ and ${\cal H}^\prime_\nu$ correspond to
two different flavor bases as generally defined in Eqs. (5) and (9),
and only in the latter basis the exact analytical results of $m^2_i$
(for $i=1,2,3$), $\theta^{}_{ij}$ (for $ij=12,13,23$) and $\delta$
obtained above are safely applicable.

We remark that $\theta^{}_{23} =\pi/4$ is strongly favored by
current neutrino oscillation data \cite{GG}. Although $\delta =\pm
\pi/2$ is purely a phenomenological conjecture, it corresponds to
maximal leptonic CP violation and thus is very interesting. In fact,
it is possible to obtain $\delta =\pm \pi/2$ from a number of
neutrino mass models \cite{Review}. If neither the $\mu$-$\tau$
flavor symmetry nor maximal CP violation is realistic, one may still
use the exact analytical relations between the elements of
$H^\prime_\nu$ and seven physical quantities to discuss a specific
neutrino mass model. To do so, however, one must carefully choose
the flavor basis of three neutrino fields so as to eliminate or
factor out the relevant unphysical phases hidden in the original
Majorana neutrino mass matrix.

Of course, one may follow the same procedure to directly diagonalize
$H^{}_\nu = V\widehat{M}^2_\nu V^\dagger = P^{}_l H^\prime_\nu
P^\dagger_l$ in a generic flavor basis. In this case the analytical
results of $m^{}_i$ (for $i=1,2,3$), $\theta^{}_{ij}$ (for
$ij=12,13,23$) and $\delta$ are more complicated and less useful
than the ones obtained above, simply because the phases of $P^{}_l$
must be involved to cancel the unphysical phases hidden in the
matrix elements of $H^{}_\nu$. Such an exercise has been done in
Ref. \cite{Baba}. Switching off the unphysical phases and taking
account of the constraint equations, we find that it is possible to
reach an agreement between the results obtained in Ref. \cite{Baba}
and ours.

Finally, we point out an immediate and interesting application of
Eq. (36) to the analysis of terrestrial matter effects on neutrino
mixing and CP violation. Assuming a constant matter density profile,
we may write out the effective Hamiltonian responsible for the
propagation of a neutrino beam in matter in the same way as that in
vacuum:
\begin{eqnarray}
{\bf H}^{}_{\rm v} & = & \frac{1}{2 \bf E} V
\begin{pmatrix} m^2_1 & 0 & 0 \cr 0 & m^2_2 & 0 \cr 0 & 0 & m^2_3
\cr \end{pmatrix} V^\dagger \; ,
\nonumber \\
{\bf H}^{}_{\rm m} & = & \frac{1}{2 \bf E} \tilde{V}
\begin{pmatrix} \tilde{m}^2_1 & 0 & 0 \cr 0 & \tilde{m}^2_2 & 0 \cr 0 &
0 & \tilde{m}^2_3 \cr \end{pmatrix} \tilde{V}^\dagger  = \frac{1}{2
\bf E} V \begin{pmatrix} m^2_1 & 0 & 0 \cr 0 & m^2_2 & 0 \cr 0 & 0 &
m^2_3 \cr \end{pmatrix} V^\dagger +
\begin{pmatrix} {\bf a} & 0 & 0 \cr 0 & 0 & 0 \cr 0 & 0 & 0 \cr
\end{pmatrix} \; ,
\end{eqnarray}
where $\bf E$ is the neutrino beam energy, ${\bf a} = \sqrt{2} \
G^{}_{\rm F} n^{}_e$ stands for the terrestrial matter effects
\cite{Wolfenstein}, $\tilde{m}^{}_i$ (for $i=1,2,3$) denote the
effective neutrino masses in matter, and $\tilde{V}$ represents the
effective neutrino mixing matrix in matter \cite{Xing00}. Given
$\theta^{}_{23} = \pi/4$ and $\delta =\pm \pi/2$ for $V$ in vacuum,
the texture of ${\bf H}^{}_{\rm v}$ must be the same as ${\cal
H}^{}_\nu$ in Eq. (36). In this case ${\bf H}^{}_{\rm m}$ takes the
same texture as ${\bf H}^{}_{\rm v}$, but its (1,1) element is
different from that of ${\bf H}^{}_{\rm v}$. This difference implies
that the constraint equation in Eq. (32) does not hold for the
matrix elements of ${\bf H}^{}_{\rm m}$, and thus we are left with
$\tilde{m}^{}_i \neq m^{}_i$, $\tilde{\theta}^{}_{12} \neq
\theta^{}_{12}$ and $\tilde{\theta}^{}_{13} \neq \theta^{}_{13}$. In
contrast, the basic texture of ${\bf H}^{}_{\rm v}$ or ${\bf
H}^{}_{\rm m}$ can be reduced to ${\cal H}^\prime_\nu$ after a
proper phase transformation, so $\tilde{\theta}^{}_{23} =
\theta^{}_{23} = \pi/4$ and $\tilde{\delta} = \delta = \pm \pi/2$
must hold. This observation is apparently consistent with the Toshev
equality $\sin 2\tilde{\theta}^{}_{23} \sin\tilde{\delta} = \sin
2\theta^{}_{23} \sin\delta$ \cite{Toshev}. Our conclusion is that
the $\mu$-$\tau$ flavor symmetry with $\theta^{}_{23} =\pi/4$ and
maximal CP violation with $\delta = \pm \pi/2$ keep unchanged when
constant matter effects are taken into account for a long-baseline
neutrino oscillation experiment. We shall explore more generic
applications of our results obtained in this paper to the
description of neutrino oscillations in matter elsewhere
\cite{Zhou}.

\vspace{0.5cm}

One of us (Z.Z.X.) would like to thank M. Yasu$\rm\grave{e}$ for
very useful comments and discussions. This work was supported in
part by the National Natural Science Foundation of China under grant
No. 10425522 and No. 10875131.

\end{document}